\documentclass[USenglish,twocolumn]{article}

\ifx\directlua\undefined\ifx\XeTeXcharclass\undefined
  \usepackage[utf8]{inputenc}                           
  \else\RequirePackage[no-math]{fontspec}[2017/03/31]\fi 
  \else\RequirePackage[no-math]{fontspec}[2017/03/31]\fi 
\usepackage[sort&compress,square,numbers]{natbib}
\usepackage[big,online]{dgruyter}

\theoremstyle{dgthm}

\theoremstyle{dgdef}

\begin{document}

\articletype{Research Article}

\author*[1]{Laura Bollmers}
\author[2]{Noah Spiegelberg}
\author[3]{Michael Rüsing}
\author[4]{Christof Eigner}
\author[5]{Laura Padberg}
\author[6]{Christine Silberhorn}
\affil[1]{Integrated Quantum Optics, Department of Physics, Institute for Photonic Quantum Systems (PhoQS), Paderborn University, 33095 Paderborn, Germany, e-mail: laura.bollmers@upb.de; https://orcid.org/0009-0006-4954-0632}
\affil[2]{Integrated Quantum Optics, Department of Physics, Paderborn University, 33095 Paderborn, Germany, e-mail: noahsp@mail.uni-paderborn.de; https://orcid.org/0009-0001-3440-4738}
\affil[3]{Integrated Quantum Optics, Department of Physics, Institute for Photonic Quantum Systems (PhoQS), Paderborn University, 33095 Paderborn, Germany, e-mail: michael.ruesing@upb.de; https://orcid.org/0000-0003-4682-4577}
\affil[4]{Institute for Photonic Quantum Systems (PhoQS), Paderborn University, 33095 Paderborn, Germany, e-mail: christof.eigner@upb.de; https://orcid.org/0000-0002-5693-3083}
\affil[5]{Integrated Quantum Optics, Department of Physics, Institute for Photonic Quantum Systems (PhoQS), Paderborn University, 33095 Paderborn, Germany, e-mail: laura.padberg@upb.de; https://orcid.org/0000-0002-8376-6305}
\affil[6]{Integrated Quantum Optics, Department of Physics, Institute for Photonic Quantum Systems (PhoQS), Paderborn University, 33095 Paderborn, Germany, e-mail: christine.silberhorn@upb.de; https://orcid.org/0000-0002-2349-5443}
\title{Optimizing Ultra-Long Continuous Wafer-Scale Periodic Poling in Thin-Film Lithium Niobate}
\runningtitle{Wafer-Scale Periodic Poling in TFLN}
\abstract{Periodically poled thin-film lithium niobate (TFLN) crystals are the fundamental building block for highly-efficient quantum light sources and frequency converters. The efficiency of these devices is strongly dependent on the interaction length between the light and the nonlinear material, scaling quadratically with this parameter. Nevertheless, the fabrication of long, continuously poled areas in TFLN remains challenging, the length of continuously poled areas rarely exceeds 10~mm. In this work, we demonstrate a significant progress in this field achieving the periodic poling of continuous poled areas of 70~mm length with a 3~µm poling period and a close to 50~\% duty cycle. We compare two poling electrode design approaches to fabricate long, continuous poled areas. The first approach involves the poling of a single, continuous 70~mm long electrode. The second utilize a segmented approach including the poling of more than 20 individual sections forming together a 70~mm long poling area with no stitching errors. While the continuous electrode allows for faster fabrication, the segmented approach allows to individually optimize the poling resulting in less duty cycle variation. A detailed analysis of the periodic poling results reveals that the results of both are consistent with previously reported poling outcomes for shorter devices. Thus, we demonstrate wafer-scale periodic poling exceeding chiplet-size without any loss in the periodic poling quality. Our work presents a key step towards highly-efficient, narrow-bandwidth and low-pump power nonlinear optical devices.}
\journalyear{2025}
\journalvolume{aop}

\maketitle

\section{Introduction} 

The miniaturization of optical circuits has been a rapidly evolving field of research and development for several decades, driving advancements in integrated optics and quantum photonics. As part of this progress, various materials have been explored to enhance performance and scalability. Among these materials, thin-film lithium niobate (TFLN) is particularly promising due to its maturing technology and a unique combination of electro-optic, nonlinear and acousto-optic properties, as well as the benefits of its thin-film structure enabling high mode confinement and high integration density \cite{boes2018status, weis1985lithium, zhu2021integrated, boes2023lithium}.\\
Nonlinear optics applications commonly require phase matching, due to the different phase velocities of the interacting waves. This is typically realized in TFLN by the quasi-phase matching (QPM) scheme, which can be achieved by periodic poling \cite{feng1980enhancement, armstrong1962interactions, Poberaj2012, Mackwitz2016, Chakkoria2025,Shur2015,Weiss2025}. Here, the ferroelectric domains of the crystal are inverted periodically to provide phase compensation by inversion of the second-order susceptibility, which enables efficient frequency conversion. Periodic poling is most commonly realized by electric field poling, but other domain inversion methods are possible\cite{Chakkoria2025}. The QPM scheme offers high flexibility, because phase matching and waveguide geometries can be optimized separately \cite{Chakkoria2025,Shur2015}.\\

In recent years, significant progress has been made in the periodic poling of TFLN enabling highly efficient nonlinear optical devices \cite{wang2018ultrahigh, ma2023fabrication,Chakkoria2025,zhao2020shallow,zhu2021integrated}. The required periods depend on the involved wavelengths, the design of the waveguides and properties of the film, e.g. its dopants and stoichiometry. Typical periods in TFLN are in the range of a few µm or less \cite{wang2018ultrahigh, babel2025ultrabright, mishra2022ultra}. Recent devices even demonstrated submicron poling, unlocking novel devices, such as counter-propagating parametric down conversion (PDC) \cite{sabatti2025nanodomain}. \\

Typically, most published works on the proof-of-principle fabrication of nonlinear devices on smaller chiplets of a few cm$^{2}$ area and poling length of a single cm at most \cite{wang2018ultrahigh, ma2023fabrication,Chakkoria2025,zhao2020shallow,zhu2021integrated}. Large scale application and commercialization of nonlinear optical devices based on QPM will eventually require poling on wafer-scale. In this regard, for example, wavelength-accurate, reproducible frequency converters realized via periodic poling were shown on the scale of a wafer \cite{xin2024wavelength}. Furthermore, wafer-scale periodic poling was demonstrated in terms of periodic poling of several 1~cm long sections distributed on the whole wafer \cite{chen2024wafer}.  However, the area of poling was still limited to typical chiplet sizes. In this regard, the realization of a continuously poled wafer-scale region remains elusive. \\
The ability to fabricate ultra-long periodically poled regions is vital for several advanced applications in nonlinear optics. Specifically, wafer-scale regions allow for the creation of frequency conversion devices with dramatically increased efficiency and scalability, as well as narrow bandwidth.  Additionally, such wafer-scale regions reduce the requirement for high pump power, thus minimizing thermal effects, photorefraction and noise, enhancing device stability, and enabling operation at previously inaccessible ultra-low power levels. Nevertheless, the length of periodically poled regions in TFLN are currently limited to the millimeter scale, usually in the range of less then 10~mm \cite{wang2018ultrahigh, krasnokutska2021submicron, chen2022ultra, niu2020optimizing, dong2024high,he2024heterogeneously}. The longest reported poled region measures 21 mm by our best knowledge \cite{chen2024adapted}.\\ 

Therefore, the goal of this study, is to achieve poling long continuous areas across a full 100~mm wafer. Here, we compare two different approaches to achieve continuous, wafer-scale poling. On the one hand we pole single, 70~mm long continuous electrodes, which allows us to inverted the domains on wafer-scale with a single pulse. On the other hand, continuous poling can be achieved by accurately stitching many smaller sections into one large area. Both processes come with their own advantages and challenges, which will be discussed throughout the paper.

\section{\label{sec:phase matching}Materials and Methods}

\subsection{Periodic Poling} \label{Polung}
The most common way to fabricate periodically poled structures in a ferroelectric like TFLN is electric field poling via lithographically-structured metal electrodes \cite{Chakkoria2025,zhao2020shallow,zhu2021integrated}. Here, structured electrodes define the periodic structure, i.e. where the strongest electric field is applied. An electric field above the coercive field results in a local inversion of the ferroelectric spontaneous polarization. Fabrication of periodically poled crystals is usually started from a single-domain crystal, where the crystal features the same domain orientation everywhere before poling and only certain parts are inverted during the poling process. For x-cut TFLN, the domain orientation is in-plane. Therefore, opposing finger electrodes are fabricated on the sample surface where the high-voltage is applied. \\

\begin{figure}
	\centering
	\includegraphics[width=0.95\linewidth]{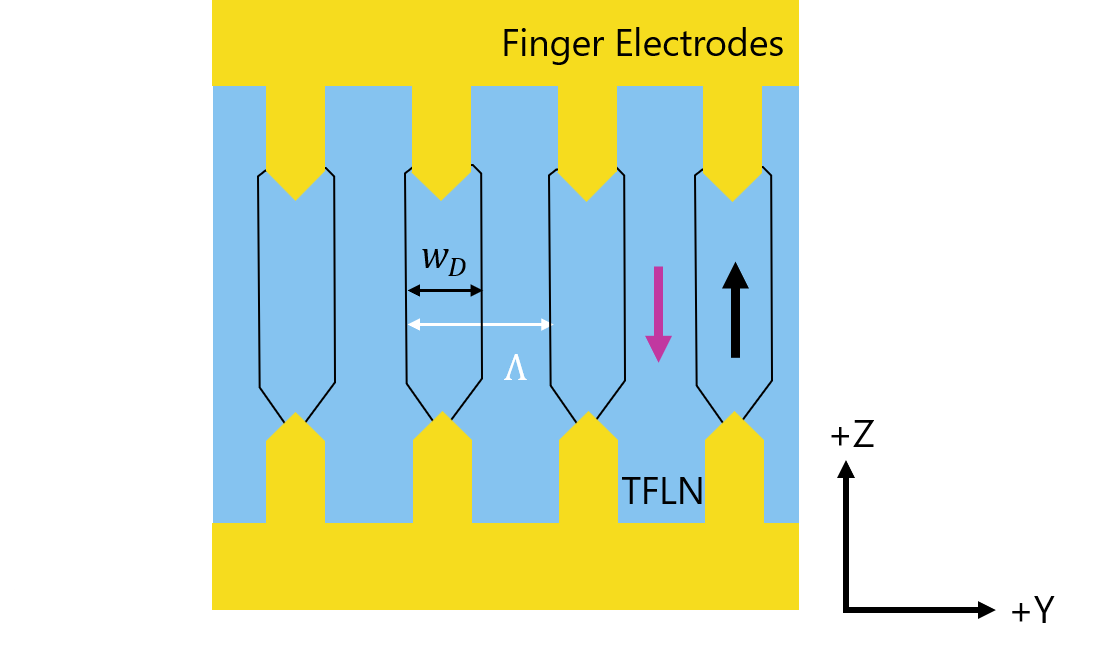}
	\caption{Schematic sketch of periodic poling via finger electrodes (yellow). Inverted domains between the opposing fingers are indicated with a black arrow. The transition, so called domain walls, between inverted and non-inverted domains (indicated by a purple arrow) are dark blue. The domains have a preferred duty cycle of 50~\%. } 
	\label{fig:SchemaPolung}
\end{figure}

For a most efficient quasi-phase matching process, the periodic poling needs to fulfill certain figures-of-merit. The first requirement is that the inverted and non-inverted domains have the same (lateral) width as it is demonstrated in Figure~\ref{fig:SchemaPolung}. In Figure~\ref{fig:SchemaPolung} a top view of a pair of finger electrodes is shown. Inverted domains are located between two opposing fingers (indicated by a black arrow, whereas the non-inverted domains are indicated by a purple arrow). The borders between inverted and non-inverted domains, so-called domain walls, appear dark blue in this sketch. Here, all domains have the same width, i.e. an ideal duty cycle. 

Poling of x-cut TFLN is usually performed with surface electrodes. This means domains have to grow into the depth of the film as well. Here, for an efficient nonlinear process it is required that the domain inversion fully penetrates through the whole thin-film thickness. Else, the optical mode in a typical waveguide will observe partially non-poled areas in the cross-section of the film, which limits conversion efficiency.
\\To quantify the poling quality in terms of the domain width, the duty cycle $\xi$ is the common figure-of-merit. The duty cycle is defined as the ratio of the width of the poled domains $w_D$, compare Figure~\ref{fig:SchemaPolung}, to the poling period $\lambda$ \cite{zhao2020shallow}:
\begin{align}
	\xi = \frac{w_D}{\lambda}
\end{align}
In an ideal case of domains with exactly the same width, the duty cycle would be 50~\%. Deviations from this as small as 10\%-points can result in a drastic drop in nonlinear efficiency, while random errors can create spectral noise and enhancement of unwanted processes especially detrimental for quantum applications \cite{Pelc2011,Zhao2023,Phillips2013,Singh2025}. To quantify the quality of a periodic poling result, it is therefore necessary to visualize the domain structure and analyze the duty cycle, as well as the depth of the domain inversion.

\subsection{Analysis of Periodically Poled Structure} \label{SHM}
The periodically poled domain structures are inspected post-fabrication with second harmonic microscopy (SHM). SHM sees widespread use for the inspection of periodically poled structures in ferroelectrics like TFLN due to its non-invasive nature, sub-micrometer resolution and its fast imaging speed \cite{Reitzig2021,Chakkoria2025,Ruesing2019,Hegarty2022,Chen2024,Floersheimer1998,Berth2009}. Here, typical microscopes enable a resolution on the order of 0.5~µm, within frames of a few hundred by a few hundred µm$^2$ in less then one second. By stitching multiple frames, large areas of cm$^2$ sizes can be inspected in reasonable times \cite{Reitzig2021,Chakkoria2025}, which makes it ideal to visualize domain structures as fabricated in this work. Specifically, in this work we use a Zeiss LSM980 microscopy with an attached femto-second light source at 920~nm for excitation. Combined with a high numerical aperture objective of 0.95,  this enables a lateral resolution of below 300~nm suitable to inspect domains structures with a 3~µm period. The SHM images are used to qualitatively inspect the poling results, as well as to quantify the duty cycle from the images and to inspect the depth of the domain inversion.

SHM relies on the effect of second harmonic generation (SHG), where two photons at the pump frequency are combined to generate a signal at twice the frequency. The effect is proportional to the local second-order nonlinear susceptibility, which is inverted when the ferroelectric domain orientation is reversed. Therefore, if the focal spot is placed on top of a domain wall, SHG light generated in one domain is out of phase by $\pi$ with respect to SHG light generated in the opposing domain. This results in destructive interference and results in a dark contrast for domain walls. At first order the domain wall can be treated to be nm-sized \cite{Gonnissen2016}, i.e. as having no substructure. Therefore this interference type contrast can be considered the dominating effect for thin films and images taken at surfaces \cite{Ruesing2019,Hegarty2022,Spychala2020}. Crucially, this interference contrast is also happening if a horizontal domain transition is present below the surface, i.e. if an inverted domain has not penetrated the full depth of film. As Rüsing et al. have shown this leads to a darker appearance of domains, which do not fully penetrate the depth of the film \cite{Ruesing2019,Reitzig2021}. As this effect is based on interference, rather then the diffraction-limited resolution, this allows to measure the penetration depth of domains in typical TFLN film with resolution down to tens of nanometer \cite{Ruesing2019,Reitzig2021}.\\

The SHM images allow not just to qualitatively inspect the poling quality, e.g. its depth or identify local errors, but to inspect the poling quantitatively. Here, by determining the location of each domain wall and measuring the width of each domain, the duty cycle can be calculated for each domain pair. From this the mean value, as well as the standard deviation can be calculated \cite{zhao2020shallow,Zhao2023,Singh2025}. This yields a measure of the absolute duty cycle, i.e. allowing to identify under- or overpoling, as well as to evaluate the homogeneity by determining the standard deviation, which is a key parameter for low-noise frequency conversion \cite{Pelc2011,Zhao2023,Phillips2013,Singh2025}.

\subsection{\label{sec:fabrication}Fabrication}
\begin{figure*}
	\centering
	\includegraphics[width=1\linewidth]{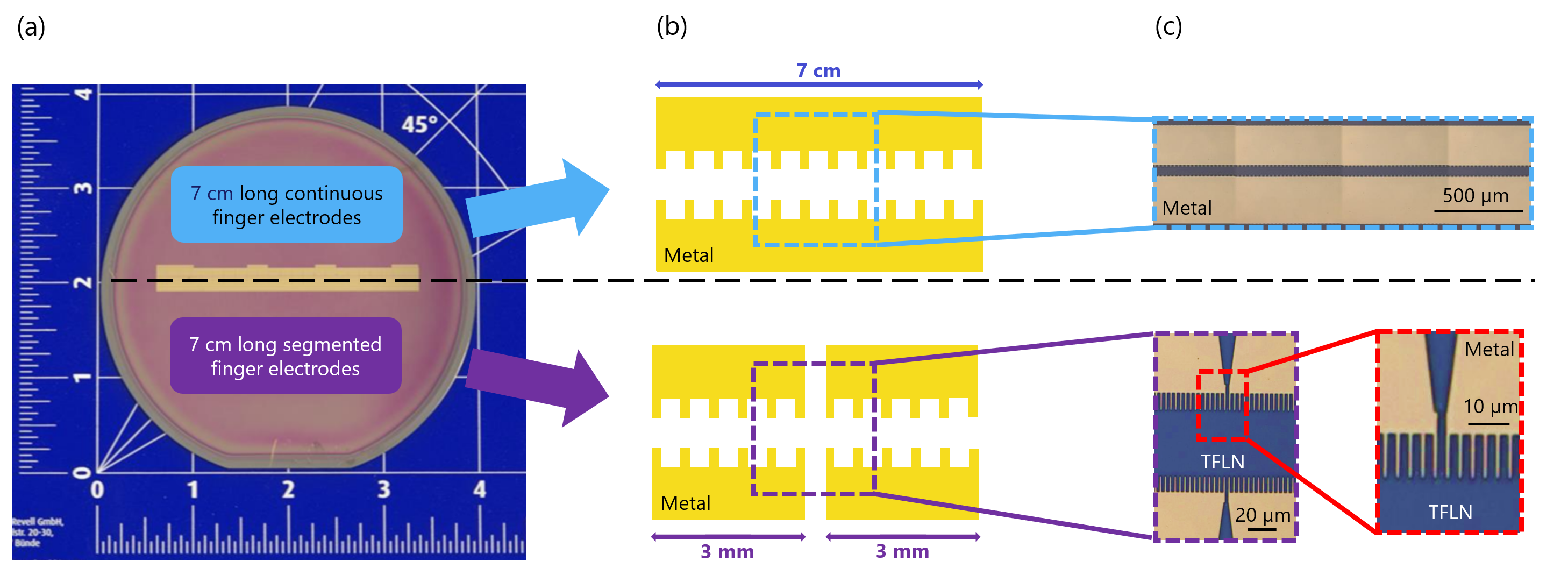}
	\caption{(a) Photograph of the wafer with fabricated finger electrodes. Here, the upper part includes 70~mm long continuous finger electrodes as well as test structures. The lower part consists of the segmented finger electrodes. A schematic sketch of the continuous and the segmented finger electrodes are demonstrated in (b). In (c) some microscope images of the two types of finger electrodes are shown.} 
	\label{fig:Elektrode}
\end{figure*}
We use a 100~mm diameter wafer (NanoLN) consisting of an x-cut 600~nm magnesium-oxide doped (5~mol$\%$) thin-film lithium niobate layer, followed by a 2~µm thick buried silicon dioxide insulator layer and a 500~µm thick silicon substrate. For fabrication we pick the central 70~mm area of the wafer, where the thickness should be most constant (thickness variations $\textless$ 20~nm according to manufacturer). In a first step, we fabricate finger electrodes consistent with a 3~µm poling period. Periods on the order of 3~µm are typical for common nonlinear-optical processes like SHG or PDC between 775~nm and the telecom band at 1550~nm \cite{zhao2020shallow,wang2018ultrahigh}. Therefore, this poling period presents a typical use-case to demonstrate poling at a wafer-scale. Please note, for the proof of principle of continuous poling we use a constant period. However, to counteract the nonlinear efficiency decrease due to thin-film variation, the poling period can be locally adapted \cite{chen2024adapted} in future work.\\
To create a continuous poling length of beyond 21~mm two possibilities exist: 1) One possible design for the finger electrodes is to extend the length of an electrode to 70~mm. This approach allows us to periodically pole along the whole wafer at once resulting in fast device fabrication. The downside of such as strategy is that the poling quality along the full electrode length can be subject to thickness or quality variations in the thin-film, such as local defects, which might impact poling. These cannot be countered by adapting the electric pulse locally. Instead, an average optimal poling pulse is required.\\
2) Another strategy is to divide the 70~mm long electrodes into smaller segments which are already known that such section can be poled with a close to 50~\% duty cycle along the full length, which is typically 3 to 5~mm \cite{zhao2020shallow}. The disadvantage of this approach is, that more electrodes need to be poled leading to an increase effort, as well as the potential negative interactions of domains during growth at the intersection of different electrodes, which might result in decreased overall quality. However, the advantage of this process is that local variations in electrode deposition or crystal quality can be optimized locally.
\\In addition to both electrode approaches for wafer-scale periodic poling, we also fabricate short finger electrode sections (0.5~mm long) on the wafer. These electrodes we use as test structure to optimize the poling pulse since every wafer is unique in its poling behavior. \\

To fabricate finger electrodes for the periodic poling process, we employ a double-resist lift-off method. For structuring we utilize electron-beam lithography (Raith Voyager) in the \textit{MBMS} (\textit{modulated beam moving stage}) mode. 
The \textit{MBMS} mode is specifically designed to create periodic patterns and is capable of avoiding stitching errors. Therefore, this mode allows us to fabricate finger electrodes with constant poling periods even for 70~mm long structures. After lithography the double-resist layer approach allows us to develop both resists individually. Hence, we are able to create an undercut resist structure which is central for a successful lift-off process. As a next step, we sputter chromium and gold on the developed sample. Here, chromium acts as the adhesion layer. To finalize the finger electrodes for subsequent periodic poling a standard lift-off process is employed.
\\In Figure~\ref{fig:Elektrode} (a) a photograph of the wafer with 70~mm long continuous finger electrodes as well as the test structures are shown (upper part). The wafer is subdivided in two parts indicated by the dashed line. The top part features continuous electrodes while the bottom features the segmented electrodes.  A schematic sketch of each area is shown in Figure~\ref{fig:Elektrode} (b), whereas zoomed in microscope pictures are shown in Figure~\ref{fig:Elektrode} (c).\\
For the second design approach, the segmented finger electrodes, we divide the 70~mm long finger electrodes into 23 shorter electrodes of approx. 3~mm length each. This approach allows us to periodically pole each segment individually. However, it is crucial to maintain electrical insulation and separation between each segment, so that neighboring sections are not influenced. Here, a white-light microscope image of such a transition is shown in Figure~\ref{fig:Elektrode} (c) (lower part). For the fabrication of the segmented finger electrodes we can still use the \textit{MBMS} mode of the electron-beam lithography. In this way we can guarantee that the period of the fingers is without stitching errors and therefore constant.
\\After the fabrication of both types of finger electrodes, we perform the periodic poling. First, we use the test structures to find an optimal poling pulse, i.e. yielding close to 50~\% duty cycle and domains penetrating the full film thickness. Subsequently, we apply a single high-voltage pulse to the larger electrodes with the optimized parameters.  For the segmented electrodes, this optimization cycle process is repeated locally by using test structures near to the segments. The poling pulse is chosen similar to Younesi et al. \cite{Younesi2021} and features a fast ramp above the coercive field, which is held for a time $\tau$ and a slower decrease of the voltage to inhibit spontaneous back poling. For a gap of 40~µm we use a voltage in the range slightly above 1.2~kV. This corresponds to an electrical field of above 30~kV/mm, which is in line with previous works on TFLN \cite{nagy2019reducing, park2022high, niu2020optimizing}.  The exact voltage required to achieve an optimal duty cycle is dependent on the position on the wafer, as discussed below.

\section{\label{sec:results}Results}

\begin{figure*}
	\centering
	\includegraphics[width=0.99\linewidth]{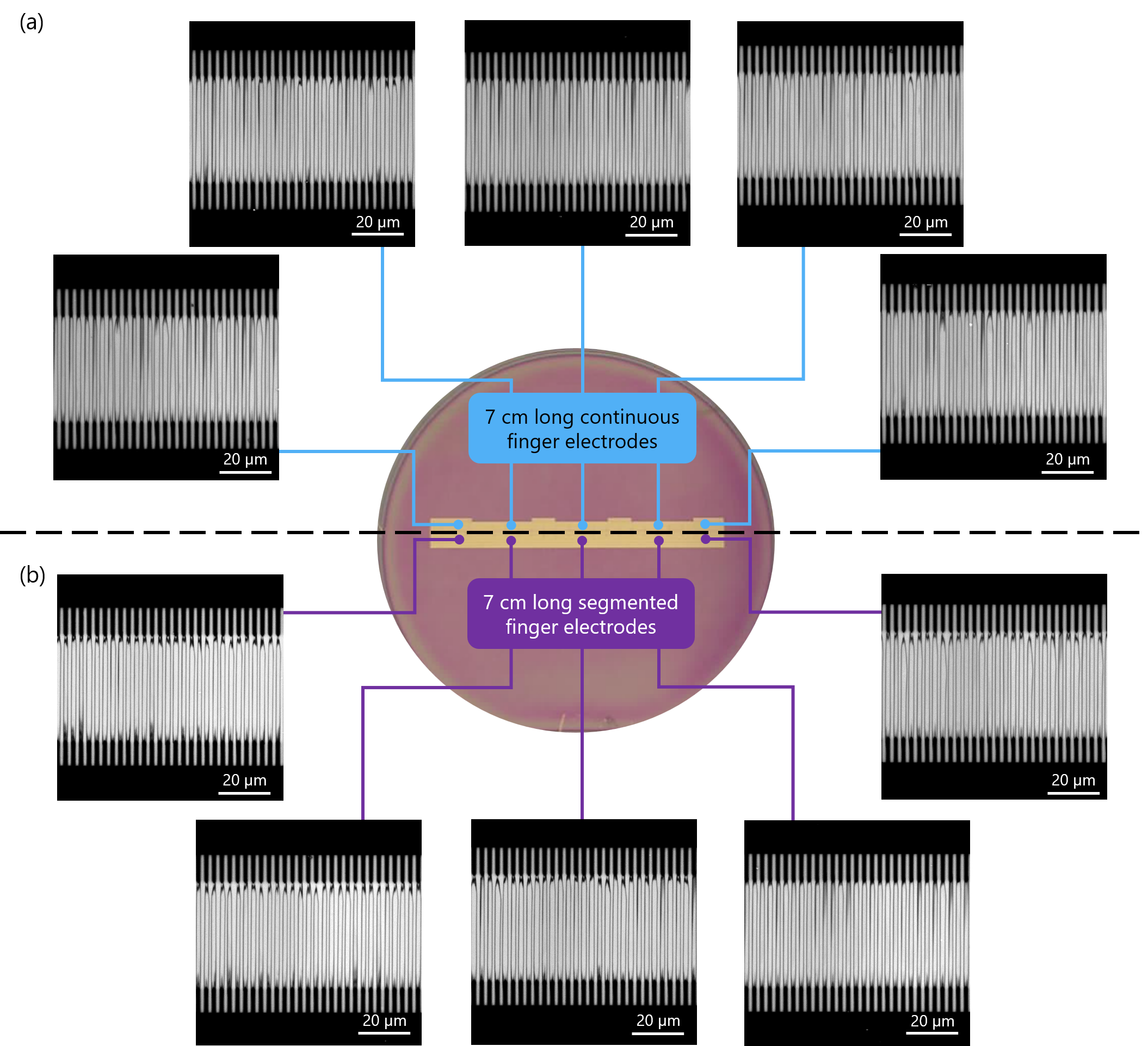}
	\caption{Second-harmonic microscope images of the periodic poling results. In the upper part (a) some examples of a 70~mm long continuous finger electrode are shown, whereas the images in (b) are taken from segmented electrodes.} 
	\label{fig:Polung}
\end{figure*}

Figure~\ref{fig:Polung} (a) shows some examples of SHG images of the periodic poling after optimization of the poling for both, the (a) continuous finger electrodes and (b) the segmented electrodes arranged around the photo of the wafer. The images are taken at different locations, respectively, and overall appear similar. The bottom and top of each SHM image shows the poling electrodes and fingers, which appear black, because the metal of the finger electrodes do not produce any SHG light. In between the opposing fingers domain walls are visible resembling the sketch in Figure~\ref{fig:SchemaPolung} very well for each location. The domain walls appear dark in the image as discussed in section \ref{SHM}. The inverted domains appear a bit wider than the non-inverted ones, which indicates a slight overpoling. For both approaches, qualitatively similar domain structures are obtained with periodic domains and no large poling defects, e.g. missing inverted domains. This first inspection reveals that, in principle, both approaches enable continuous poling of 70~mm long domain structures.\\
However, a closer look reveals differences between the poling results of the (a) continuous single finger electrode compared to (b) the segmented finger electrodes. Most notably, the domains from the continuous electrode appear less homogeneous in their width compared to the results from the segmented electrodes (b). This indicates on average a larger random duty cycle error for the continuous electrode, which could be detrimental for certain nonlinear applications \cite{Pelc2011,Zhao2023,Phillips2013,Singh2025}. For the segmented electrodes the domain width of each domain appears to be more consistent and similar.\\

\begin{figure*}
	\centering
	\includegraphics[width=0.9\linewidth]{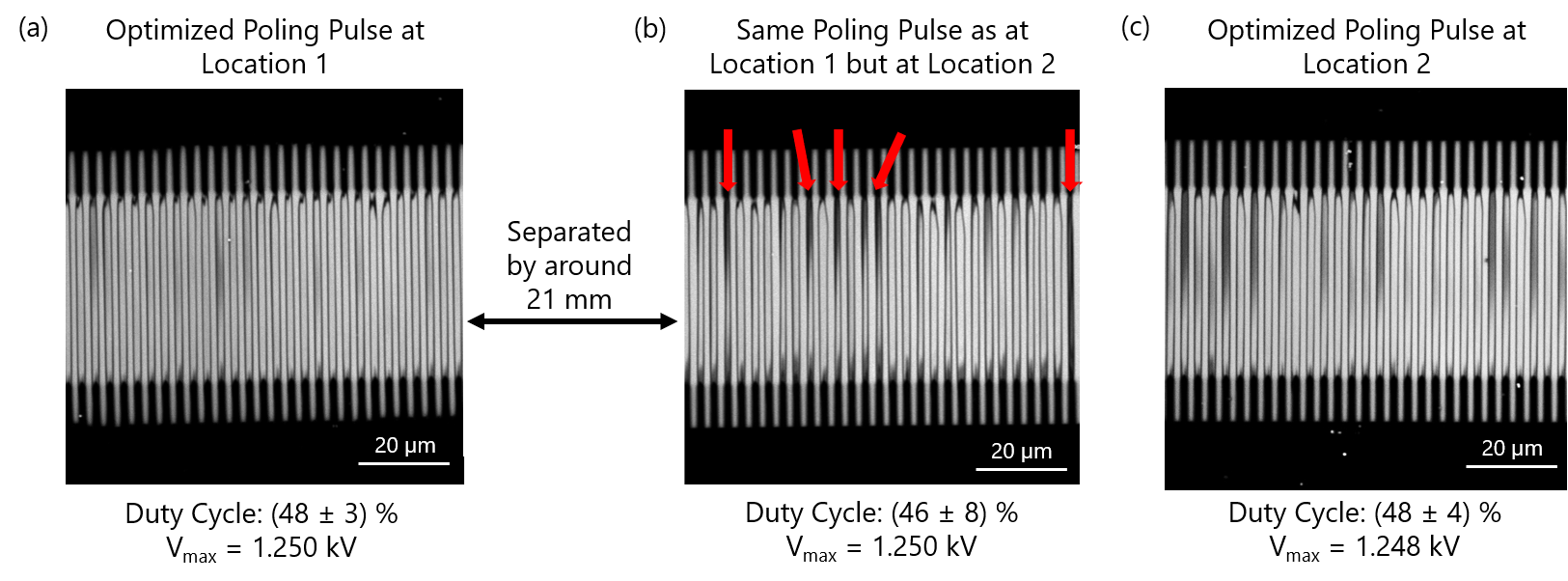}
	\caption{SHM image of three segmented electrodes. To two electrodes the same poling pulse were applied. These electrodes are separated by about 21~mm along the y-direction of the wafer, but are otherwise equal. The poling pulse optimized for electrode at location 1 (a) was used unaltered for location 2 (b). The periodic poling result in (b) is not as good as the one in (a) since the deviation in the domain width increases as well as some dark domains (highlighting with red arrows) appear indicating a surface-near poling. This highlights the need for local optimization of poling pulses as it is done in (c). Only a small variation in the applied voltage, $V_{max}$, yields a significant improvement in the resulting homogeneity and therefore of the overall duty cycle.} 
	\label{fig:gleicherPuls}
\end{figure*}

A possible explanation for the more inhomogeneous poling result is the fact poling a 70~mm long stretch with a singular poling pulse is challenging, because no single poling pulse is ideal for the full length of a continuous electrode. Here, local variations in crystal stoichiometry or electrode homogeneity lead to slightly different required poling pulse for an optimized poling result. To illustrate this, we can compare the poling results of two segmented electrodes approximately 21~mm apart along the length of the electrode. Such an example is shown in Figure~\ref{fig:gleicherPuls}. Here, a poling pulse was optimized for the first location by poling several 0.5~mm long test sections in the direct vicinity. The result is a poling pulse, which yields an overall close to 50\% duty cycle with a very small deviation (48 $\pm$ 3)~\% on the test structure, see Figure~\ref{fig:gleicherPuls} (a). If the same poling pulse without further change is applied to another test structure at a second location about 21~mm along the y-direction, then a more inhomogeneous domain structure is obtained with almost twice the deviation (46 $\pm$ 8)~\%. This is also seen in features of the image, such as randomly appearing narrow and dark domains, which are highlighted with red arrows in Figure~\ref{fig:gleicherPuls} (b). Dark domains can be related to surface-near poling meaning the inverted domains have not grown deep into the thin-film. If the poling pulse is optimized again at the new location then an improved homogeneity and duty cycle closer to 50\% can be achieved, see Figure~\ref{fig:gleicherPuls} (c). This can be obtained by only a small change of the poling pulse. Thus, we reduced the applied voltage, $V_{max}$ by 2~V which corresponds to a change less than 0.2~\%. Both, the overall duty cycle as well as its deviation is improved to (48 $\pm$ 4)~\% due to this small adjustment of the poling pulse. It is noteworthy that this local change in the poling behavior occurs on that scale which was up to now the maximum length of a periodically poled area (21~mm). This highlights the challenging nature of wafer-scale periodic poling in TFLN.

To counteract this problem, we can utilize the second approach of segmented finger electrodes with locally optimized poling pulses. Examples of the corresponding poling results are shown in Figure~\ref{fig:Polung} (b). The SHM images show more homogeneous poling results. Before we compare the poling results of both approaches, we focus on the behavior exhibited at the transitions between adjacent segments in the segmented configuration.
\begin{figure*}
	\centering
	\includegraphics[width=0.83\linewidth]{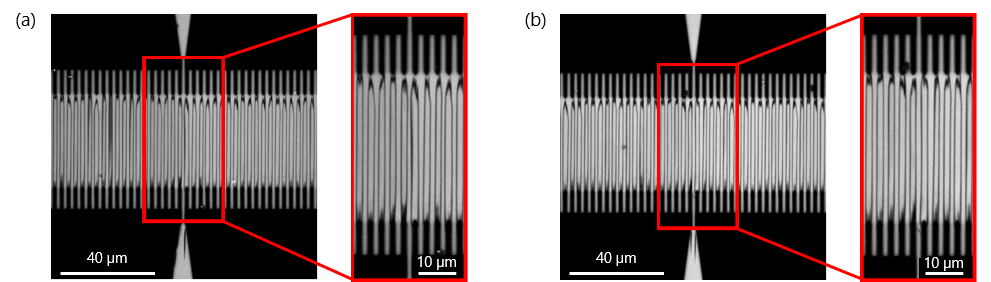}
	\caption{SHM image of segmented finger electrodes with a close-up to the transition. Neither in (a) nor in (b) no significant difference can be seen in the poling quality at the transitions of two segments compared to the periodic poling results near the transition.} 
	\label{fig:Segmentierung}
\end{figure*}
Here, already poled domains in one segment might influence or inhibit growth of domains or electrostatic induction result poling or influence in neighboring segments. Figure~\ref{fig:Segmentierung} shows SHM images of two examples including close-ups of the transition region. Here, no significant differences in poling quality between the transitions or other regions can be detected. Indeed, a closer look at the close up of Figure~\ref{fig:Segmentierung}(a) reveals that the inverted domains at the transition are slightly closer together than the others but have not even merged. On the length of 70~mm ($>$ 23000 periods) these transitions account for only 0.1~\% of all periods. Thus, this potential weak point in the segmented electrodes approach has no significant impact on the poling results.

\section{\label{sec:discussion}Discussion}
\begin{figure*}
	\centering
	\includegraphics[width=1\linewidth]{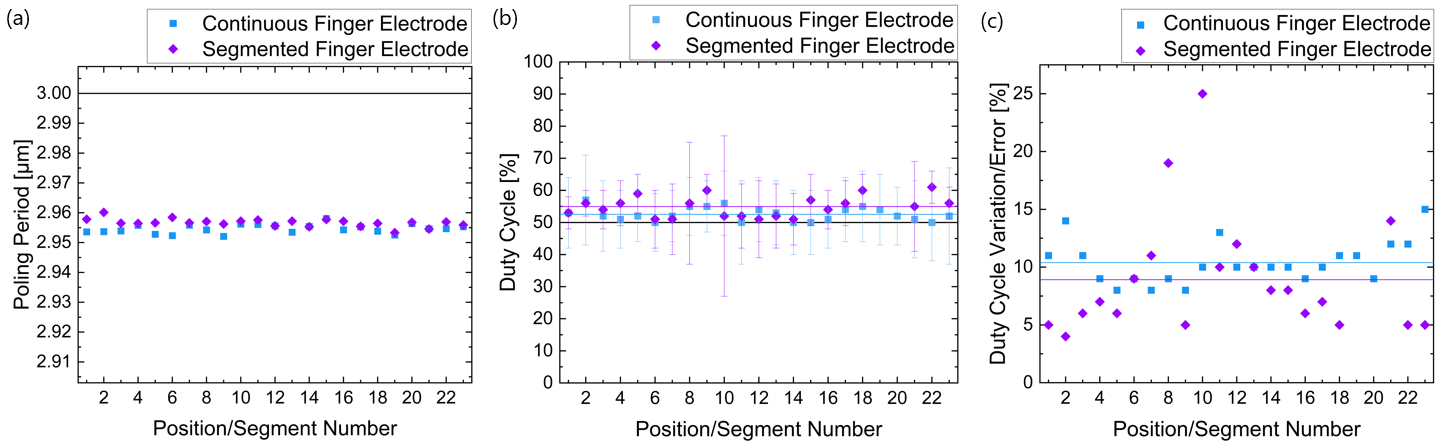}
	\caption{Comparison of the poling periods, the duty cycles and their errors of the two periodic poling approaches: (a) the measured poling periods for the continuous (blue) and the segmented finger electrodes (purple) do not differ from each other as expected. The constant shift is in the range of 1.5~\% and probably due to some calibration offset. (b) duty cycles of the periodic poling results for the continuous (blue) and the segmented finger electrodes (purple), the corresponding lines indicates the mean values and the black one the targeted duty cycle. The poling results from the continuous electrode is divided into smaller parts to compare it with the segmented one. The lines represent the arithmetic mean of the duty cycle along the whole 70~mm long periodically poled area for both approaches.(c) error distribution of the duty cycle for the continuous (blue) and the segmented finger electrodes (purple) approach.  The lines represent the (arithmetic) mean of the duty cycle deviation taking of all electrode segments respectively positions.} 
	\label{fig:DC}
\end{figure*}

As a next step, we compare the periodic poling results systematically by determining the poling period, the duty cycle and its deviation for both approaches. For this purpose we divide periodically poled 70~mm long areas into smaller parts. Since the segmented finger electrode is already subdivided into 23 segments, we chose the same number of parts for the continuous finger electrode. This allows us to compare the poling results of the finger electrode segments with a similar position on the continuous electrode. For each of these locations we take SHG images and extract the poling period and duty cycle. A single frame taken with the SHM has a size of 70~µm $\times$ 70~µm. Thus, we stitch for each location several frames together to get an image of a full segment. To keep the analysis consistent and comparable we analyze for each segment at least 100 periods at random locations to minimize the number and therefore the impact of stitching error which occur in the SHM. For each segment we repeated this process multiple times at different random locations and did obtain the same results within the uncertainties every time. The results are shown in Figure~\ref{fig:DC}. \\
The calculated poling periods of every segment/position do not differ from each other whether they belong to the continuous (blue) or segmented finger electrodes (purple). The poling periods are demonstrated in Figure~\ref{fig:DC} (a). Here, the black line is the reference from the set poling period. The measured poling period deviates by approx. 0.05~µm, which is a deviation in the range of 1.5~\%. The constant shift from the targeted poling period is probably due to some calibration offset. Nevertheless, we established a fabrication process for finger electrodes which allows for the realization of a constant poling period even for wafer-scale.\\
The absolute values of the duty cycle for the continuous (blue) and segmented finger electrodes (purple) in Figure~\ref{fig:DC} (b) are close to 50~$\%$. The arithmetic mean of the duty cycle along the whole 70~mm are represented by a blue line (continuous finger electrode) and a purple one (segmented finger electrode) in contrast to the black line which represents the targeted 50~\% duty cycle. In this regard a corridor of $\pm$10~\% around 50~\% is generally considered acceptable for efficient nonlinear devices \cite{Pelc2011,Zhao2023,Phillips2013,Singh2025}. However, there are significant differences in the deviation. Segments number 8 and 10 of the segmented finger electrodes exhibit notably larger errors than the others. This indicates a key challenge with segmented finger electrodes: achieving consistent, high-quality poling across all segments. This also can be seen in the fact that we were not able to calculate a duty cycle for the segments number 19 and 20, since the inverted domains did not completely grow into the thin-film and therefore appear completely dark in our experiment. These challenges with the segmented approach might be overcome with the use of further test structures or applying further poling pulses to improve the results \cite{Rao2019,Zhao2019}. Further, while these segments (partly) failed, it should be noted that poling of the continuous 70~mm electrode also is a hit-or-miss situation, but with only a \emph{single} chance. If poling the single electrode fails, the full structure is worthless. In contrast, if a few, individual sections are not poled or are drastically over- or underpoled they have much less impact on the total efficiency and spectrum of a possible nonlinear optical device, as they only make up a small portion of the total structure and the nonlinear efficiency, which even scales quadratically with total interaction length. Nevertheless, beside the large errors for the segments number 8 and 10, Figure~\ref{fig:DC} (b) demonstrates no significant difference in terms of absolute duty-cycle between the two design approaches.\\

The key difference between the two approaches can be better seen, if the standard deviations are inspected separately from the absolute duty cycle. For this, the standard deviations are plotted in Figure~\ref{fig:gleicherPuls} (c). Here, it is apparent that with the exception of the outliers of segments 8 and 10, almost all standard deviations of the segmented approach are often below 5~\%, while the errors for the sections of the continuous electrodes are in the order of 10~\%. If we calculate the (arithmetic) mean duty cycle deviation for both (blue and purple line), we get a value of (10.4 $\pm$ 1.9)~\% for the continuous and (8.9 $\pm$ 5.2)~\% for the segmented finger electrode. Here, even though the strong deviations of segment 8 and 10 are taken into account, the average deviation is still smaller for the segmented electrodes. Without taken segments 8 and 10 into account, the value falls as low as (7.5 $\pm$ 2.8)~\%. Further inspecting the deviations, it appears that the continuous approach also results in stronger deviations towards the ends of the electrodes, where the deviation rises above 10~\% towards both ends. This might be a result of a voltage drop-off toward the edges of the electrodes. In contrast, the segmented finger electrodes do not show this trend. Here, the variation of the error is distributed more randomly.\\

For quantum applications, structures with such low deviations are particularly desirable \cite{Pelc2011,Zhao2023,Phillips2013,Singh2025}. The ability to minimize random duty cycle errors helps reduce spectral noise and enhances nonlinear processes, which is crucial for quantum optical devices. Although the mean duty cycle of the segmented electrode structure deviates slightly more from the target 50~\% compared to the continuous electrode, both results are consistent with previously reported poling outcomes for shorter devices. For instance, Dong et al. demonstrated a 4~mm long TFLN waveguide with calculated duty cycles of (48~$\pm$~4)~\% and (42~$\pm$~5)~\% \cite{dong2024high}. Please note, that they calculated their duty cycles for a 30~µm long poled section. Moreover, duty cycles differing by 5~\% or less from the ideal 50~\% still yield over 90~\% of the conversion efficiency of an ideal QPM process \cite{wagner2008modeling}.\\
These findings indicate that both approaches allow not only wafer-scale periodic poling but also maintain duty cycles comparable to those observed in shorter structures. From the perspective of periodic poling, this enables the extension of device lengths beyond the typical 1–2~cm, reaching wafer-scale dimensions. By employing segmented electrodes, structures longer than 70~mm can be poled without additional process adjustments, as only the number of segments needs to be increased. Furthermore, with additional test structures, the poling pulse can be individually optimized for each segment, offering the potential to further improve poling quality which is highly interesting for quantum optical application. Consequently, periodically poled structures are no longer restricted to chiplet-scale device without compromising poling quality.

\section{Conclusion}
In this paper, we demonstrate high quality wafer-scale periodic poling of TFLN over 70~mm, achieving duty cycles close to  50~\%. Since the conversion efficiency of devices scales quadratically with the interaction length and therefore with the length of the periodically poled area, devices for high-power frequency comb generation benefits strongly from ultra-long poled structures. Both, continuous and segmented electrode designs, yield results comparable to those observed in much shorter devices, confirming that poling quality is maintained even at wafer-scale lengths. Continuous electrodes offer the advantage of fast, high-throughput fabrication, making them ideal for large-scale production. In contrast, segmented electrodes provide enhanced flexibility for implementing complex or aperiodic poling patterns, enabling broadband and tailored nonlinear interactions. Furthermore, by optimizing the poling pulse for each segment individually, random duty cycle errors can be minimized, which is particularly advantageous for high-performance quantum frequency converters requiring low spectral noise. Overall, the successful demonstration of wafer-scale, high-quality periodically poled TFLN represents a major step toward scalable nonlinear and quantum photonic devices, bridging the gap between chiplet-scale prototypes and practical large-scale implementations.

\begin{funding}
The Deutsche Forschungsgemeinschaft (DFG, German Research Foundation) – SFB-Geschäftszeichen TRR142/3-2022 – Projektnummer 231447078
\end{funding}

\begin{authorcontributions}
L.B. and L.P. developed the idea. L.B. performed the experiment including the fabrication of the electrodes, the periodic poling and took the SHM images. N.S. and M.R. realized the code to calculate the duty cycle and poling period, where L.B. analyzed the data. L.B. and M.R. wrote the manuscript with contributions from C.E., L.P. and C.S.. The project was supervised by L.P. and C.E.. C.S. initiated the work. All authors discussed the results and provided feedback to the manuscript. All authors have accepted responsibility for the entire content of this manuscript and approved its submission.


\end{authorcontributions}

\begin{conflictofinterest}
Authors state no conflict of interest.
\end{conflictofinterest}

\begin{dataavailabilitystatement}


The datasets generated during and analyzed during the current study are available in the ZENODO repository, https://doi.org/10.5281/zenodo.17092635.

\end{dataavailabilitystatement}


\end{document}